\documentclass[english, aps, preprint]{revtex4-1}
\usepackage[T1]{fontenc}
\usepackage[latin9]{inputenc}
\usepackage{color}
\usepackage{babel}
\usepackage{verbatim}
\usepackage{graphicx}
\usepackage[unicode=true, pdfusetitle, bookmarks=true, bookmarksnumbered=false, bookmarksopen=false, breaklinks=false, pdfborder={0 0 1}, backref=false, colorlinks=true]{hyperref}
\hypersetup{linkcolor = blue,  urlcolor  = blue, citecolor = blue, anchorcolor = blue}
\usepackage{breakurl}

\newcommand{\vm}{v_M}
\newcommand{\Ma}{{\rm Ma}}

\begin{document}

\title{Marangoni elasticity of flowing soap films}
\author{Ildoo Kim}
\email{ildoo_kim@brown.edu}
\affiliation{School of Engineering, Brown University, Providence, Rhode Island 02912}
\author{Shreyas Mandre}
\affiliation{School of Engineering, Brown University, Providence, Rhode Island 02912}

\begin{abstract}
We measure the Marangoni elasticity of a flowing soap film to be 22 dyne/cm irrespective of its width, thickness, flow speed, or the bulk soap concentration. 
We perform this measurement by generating an oblique shock in the soap film and measuring the shock angle, flow speed and thickness.
We postulate that the elasticity is constant because the film surface is crowded with soap molecules.
Our method allows non-destructive measurement of flowing soap film elasticity, and the value 22 dyne/cm is likely applicable to other similarly constructed flowing soap films. 
\end{abstract}

\date{\today}

\maketitle


Stationary and flowing soap films are an ideal experimental device to simulate two-dimensional (2D) flows.
The development of a soap film channel as a scientific instrument\cite{Gharib:1989al,Chomaz:1990vl,Beizaie:1997to,Rutgers:2001wj, Georgiev:2002kg} expedited the exploration of fundamental physical fluid dynamics problems using 2D hydrodynamics.
Applications include investigations of cylinder wake\cite{Vorobieff:1999dd,Vorobieff:1999wn,Roushan:2005un,Kim:2015jp}, the flow past flapping flags\cite{Jung:2006tn,Ristroph:2008wh}, 2D decaying and forced turbulence\cite{Kellay1995, Martin:1998ty,Rivera:1998tw,Rutgers:1998uz,Jun:2005td} and 2D pipe flow\cite{Tran:2010hn}.

The persistence of a freely suspended soap film and its mechanical stability is because of soap molecules acting as surfactants\cite{Couder1989,Chomaz:2001us}.
The Marangoni effect arising from the surfactant imparts an elasticity $E=2A~(d\sigma/dA)$ to the soap film; an increase in the area $A$ of a patch of the film, which necessarily accompanies film thinning, causes the surfactant molecules to spread apart and the surface tension $\sigma$ to increase.
This increase provides a restoring force that tends to dynamically recover the original area of the film. 
In this manner, the same mechanism that stabilizes the soap film also imparts a compressible character to the 2D flow in the film. 
This compressible character is integral to, and therefore an unavoidable consequence of, the mechanism that stabilizes the film.
The degree of compressibility is quantified by comparing the characteristic soap film flow speed $u$ with the Marangoni wave speed $\vm = \sqrt{2E/\rho h}$, where $\rho$ is the fluid density and $h$ is its thickness \cite{Taylor:1959ta,Couder1989,Chomaz:2001us}.
If $\Ma = u/\vm \ll 1$, then the inertial forces in the film are too weak to overcome the elastic forces and the film is assumed to approach incompressibility\footnote{The compressibility is proportional to the square of Mach number like gas flows. Using the momentum equation for a steady soap film $\rho u d u   =  d \sigma/ h\label{eq:momentum}$ and the definition of elasticity $d\sigma = -E dh/2h = - \rho \vm ^2 dh/4$, yields $(dh/h) = - 4 \Ma^2 (du/u)$.}\cite{Auliel:2015dm}.

The objective of this paper is two-fold; to present a simple method to measure the Marangoni wave speed and to use the measured wave speed to characterize the film elasticity.
Despite the widespread use of soap films for simulating 2D fluid system, $\vm$ or $E$ is not typically measured or reported.
It is desirable to monitor $\vm$ given that it may change with the operational parameters of the soap film and that it could be comparable to the typical velocity scale of the simulated 2D flows.
Indeed, based on separate measurements of $\vm$ (246 to 362 cm/s \cite{Wen:2003wz}) and the typical flow speeds (150 to 250 cm/s\cite{Jung:2006tn}, 100 to 400 cm/s\cite{Tran:2009vf} or 270 to 600 cm/s\cite{Fayed:2011iw}), the soap film flows may not be assumed to be incompressible.
However, techniques presented in the literature to measure $\vm$\cite{Prins:1967tk,Wen:2003wz} are too cumbersome to be adopted for repeated real-time monitoring.

We present a simple technique based on an analogy with compressible gas dynamics \cite{Wen:2001vf,Wen:2003td,Wen:2003wz,Wen:2004uv,Fast:2005wq,Wen:2008vo} to measure the Marangoni wave speed. 
Our technique involves inserting a thin cylinder (a needle) in the soap film, and if required dragging it through, to generate an oblique shock.
Then shock angle $\beta$ is used to determine $\Ma$ of the incoming flow.
Dragging the needle through the film increases the relative speed and allows the shock to form even when the soap film flow is subcritical, which is analogous to {\it subsonic} flow.
We have found that, with some practice, the needle can be dragged with bare hands, therefore no additional experimental setup is required.

\begin{figure}
\begin{centering}
\includegraphics[width=8.5cm]{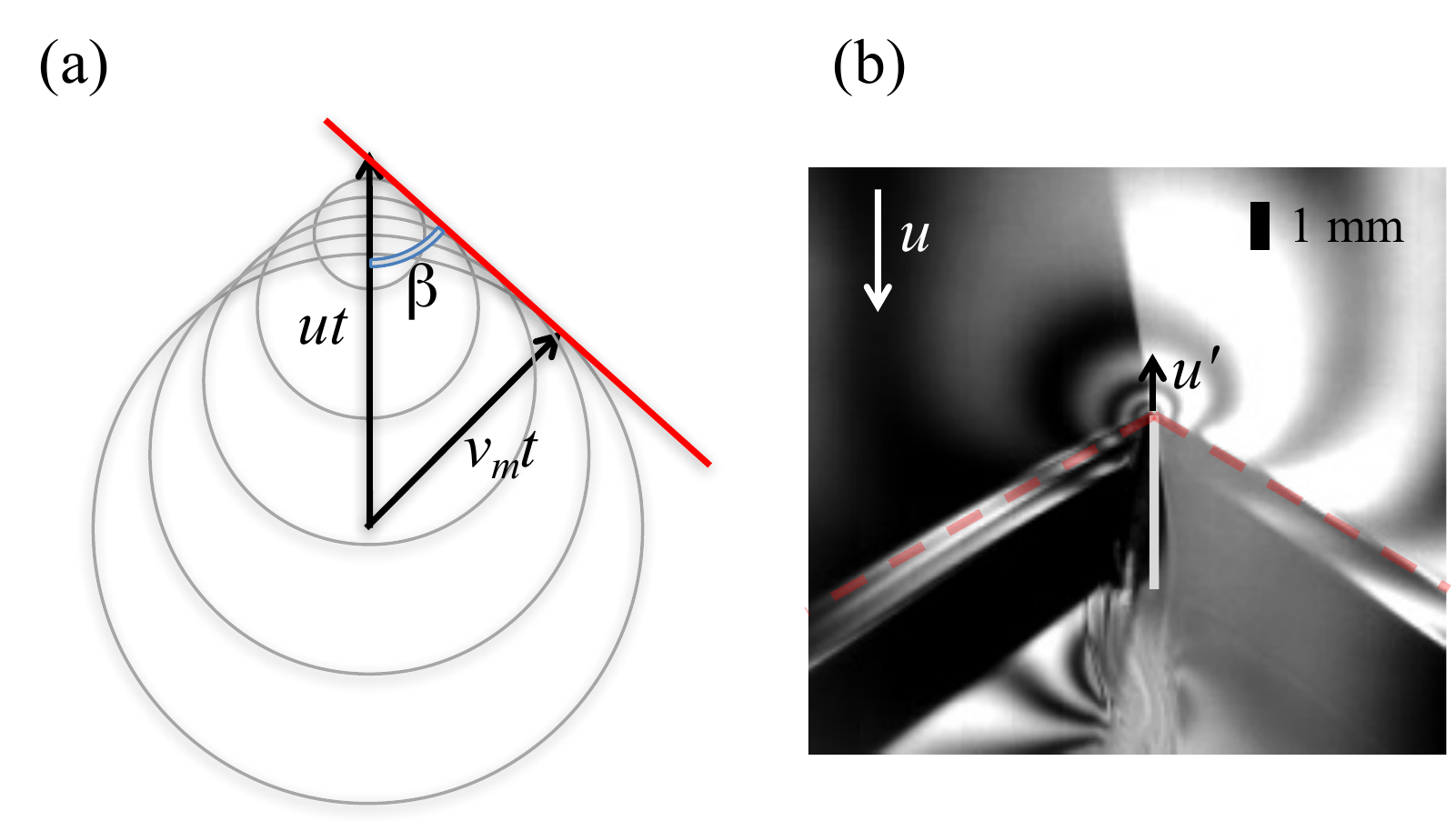}
\par\end{centering}
\caption{
(Color online)
(a) 
If a source of the wave is moved by $ut$ for a time interval $t$ and the wave is expanded by $\vm t$ for the same interval, a simple trigonometry shows the relation $\sin\beta={\vm}/{u}$. 
(b)
Typical oblique shock in a soap film flow.
To enhance experimental range, the thin plate is moved at $u'$ in a flow of speed $u$.
Shock is formed at a sharply defined angle (the dashed line serves as a guide to the eye).
\label{fig:Oblique-shock-relation}}
\end{figure}

The relationship between $\beta$ and $\Ma$ is derived from a simple geometric construction.
The shock is formed by the envelope of circular wavefronts, that originate at the obstacle, are advected by the free stream $u$ and expand at speed $\vm$ (see Fig \ref{fig:Oblique-shock-relation}(a)); this simple construction leads to the relation
\begin{equation}
\sin \beta = \frac{1}{\Ma} = \frac{\vm}{u}.\label{eq:MachConeAngle}
\end{equation}
Thus, $\Ma$ may be estimated by measuring the oblique shock angle $\beta$, and an independent measurement of $u$ yields $\vm$.
Eq. (\ref{eq:MachConeAngle}) is the special case of a more general $\alpha-\beta-\Ma$ relation for oblique shocks formed around wedges of angle $\alpha$\cite{Liepmann57}.
According to the gas dynamic analogy, the surface tension acts analogous to gas pressure, the film thickness is analogous to gas density, and the ratio $\Ma$ plays a role identical to that played by the Mach number in compressible gas dynamics; then the soap film flow is shown to be analogous to gas flow with the heat capacity ratio $\gamma=1$.
The corresponding oblique shock relation due to a wedge is presented in a previous study\cite{Wen:2004uv}.

We also present the Marangoni wave speed measured for soap films created using the commonly used solutions of commercial detergent.
We find that the Marangoni wave speed is between 330 and 200 cm/s as the film thickness varies from 4 to 11 $\mu$m.

Our measurement of $\vm$ allows us to conveniently probe the elasticity of soap films; we find that in our setup the soap film elasticity remains constant at $E=22 \,\rm dyne/cm$. 
We propose that the constant value of the elasticity is due to the overcrowding of soap molecules on the film surface. 

Our soap film channel setup is similar to those previously used by various groups\cite{Rutgers:2001wj,Martin:1998ty,Tran:2009vf,Kim:2010vt}.
The channel is vertical and is approximately 1.8 m long and 3 to 6 cm wide depending on experimental conditions.
We use 2\% solution of commercial dish soap (Dawn, P\&G) in distilled water to form soap films. The soap solution flux $F$ is controlled by adjusting the valve opening at the top of the channel.
The valve opening is calibrated to $F$ by directly collecting the solution per unit time and weighing them, and it gives the measurement of $F$ up to 5\% of uncertainty.
We work in a section of the soap film where the thickness does not vary with downstream distance, far from any ``hydraulic jump'' that can form near the end of the channel\cite{Tran:2009vf}.
The flow speed $u$ of the channel is determined by particle tracking using high speed imaging (Photron SA4).
Our measurements of $u$ are compared to separate particle image velocimetry (PIV), and the two measurements agree within 1\%.
Once $F$, the width $W$ and $u$ are determined, we can calculate the thickness of the film using $h=q/u$, where $q\equiv F/W$ is the flux per unit width.
We take all of our measurements at the center of the channel, although our method in principle can be used at any part of the soap film channel, provided that $h$ is considered as local thickness where we measure $u$.
Our separate measurements of $h(x)$ using low pressure sodium lamp interferogram and $u(x)$ using PIV, where $x$ is the span-wise coordinate of the channel, reveal that $q=h(x)u(x)$ is independent of $x$.
This is also indicated by other studies\cite{Tran:2009vf, Rutgers:1996vg}.

In the usual experimental conditions, $q$ is varied from 0.1 $\rm{cm^2/s}$ to 0.4 $\rm{cm^2/s}$.
Under such conditions, $u$ varies from 250 to 330 cm/s, and $h$ varies from 4 to 11 $\mu\rm{m}$. 
A simple dimensional analysis to balance the gravitational force and air friction implies $u\sim q^{2/5}$ and $h\sim q^{3/5}$\cite{Rutgers:1996vg}, and this is roughly consistent with our observations.

To simulate a wedge of $\alpha=0$, we place a thin plate in the middle of the soap film. 
The thin plate is 0.4 cm long in longitudinal direction and 25 $\rm \mu m$ wide in thickness.
If the flow speed is greater than the Marangoni wave speed, namely $u>v_M$, an oblique shock is formed on both sides of the $\alpha=0$ wedge.
Otherwise, when $u<v_M$, no shock is observed; we then move the thin plate against the soap film flow along a translational stage at the speed $u'$ in the lab frame.
A simple Galilean transformation gives the relative speed $v$ between the flow and the wedge as $v=u+u'$.
This technique grants us two important features: to observe oblique shocks when the flow is naturally subcritical, and to achieve a greater range of $\Ma$.

We note that our results are reproducible when the $\alpha=0$ wedge is replaced by a thin needle.
Unlike a thin plate shock, a thin needle shock is insensitive to the angle of attack. 
Therefore our scheme can be adopted at no cost, even without a translational stage.



Figure \ref{fig:Oblique-shock-relation}(b) shows a typical oblique shock formed at an angle $\beta$ relative the $\alpha=0$ wedge inserted in a soap film channel.
Analyses of images like the one in Fig. \ref{fig:Oblique-shock-relation}(b) give the measurements of $\beta$ as a function of flow conditions.
To reduce the uncertainty, the measurements are repeated six times per flow condition.

\begin{figure}
\begin{centering}
\includegraphics[viewport=50bp 20bp 680bp 520bp,clip,width=3.1in]{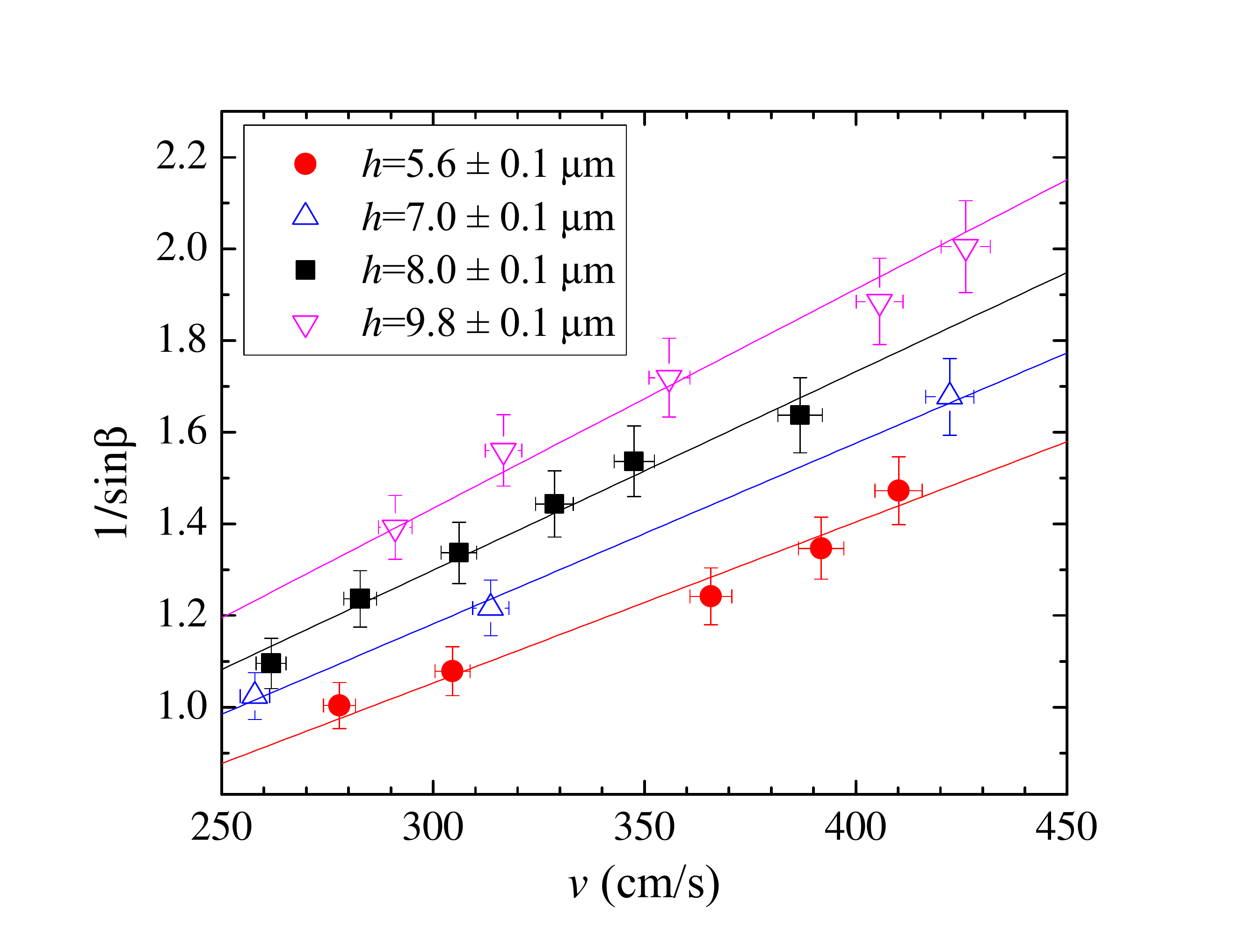}
\par\end{centering}
\caption{
(Color online) 
($1/\sin\beta$) vs. $v$ at different film thicknesses. 
Eq. (\ref{eq:MachConeAngle}) suggests that these two quantities are linearly proportional to each other when the Marangoni wave speed $v_M$ is constant.
Such linearity is observed when experimental data points are grouped by the film thickness $h$.
For a fixed $h$, ($1/\sin\beta$) is directly proportional to $v$ with zero intercept, and the slope decreases as $h$ decreases.
This relationship implies that the Marangoni wave speed increases as the film gets thinner.
\label{fig:firstfigure}}
\end{figure}

In the prescribed setup, the gasdynamic analogy in Eq. (\ref{eq:MachConeAngle}) implies a linear relation between $(1/\sin\beta)$ and the relative speed $v$, and we find that the linearity is observed only when we group data by their corresponding film thickness.
For example, data with $h=5.6 \pm 0.1 \rm \mu m$ are grouped together and displayed in Fig.  \ref{fig:firstfigure} as circles.
As the corresponding solid line indicates, $(1/\sin\beta)$ and $v$ are linearly proportional to each other.
By grouping similar cases for other film thicknesses, we find that $v_M$ is faster in thin films than in thick films.
Fig. \ref{fig:firstfigure} also shows experimental data for $h$=7.0,  8.0 and  9.8   $\rm \mu m$.
Here we find that for all cases the intercept is zero as expected, but the slope varies by the thickness.
The slope is the most gradual for the thinnest film (see circles in Fig. \ref{fig:firstfigure}) and the steepest for the thickest film (upside-down triangles).
In our model, the slope is reciprocal to $\vm$, therefore our observation implies that $v_M$ is faster in thinner films than in thicker films.

\begin{figure}
\begin{centering}
\includegraphics[viewport=50bp 70bp 670bp 520bp,clip,width=3.3in]{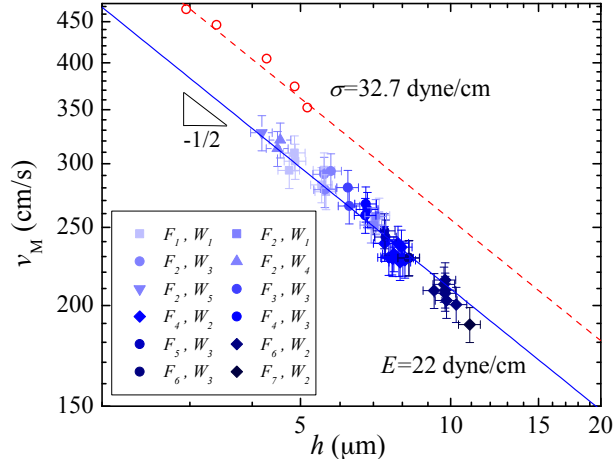}
\par\end{centering}
\caption{
(Color Online)
The Marangoni wave speed $v_{M}$ vs. the film thickness $h$ (closed symbols). 
The solid (blue) line shows $v_{M}\sim h^{-0.5}$ corresponding to $E=22 \,\rm dyne/cm$. 
The symbol and color stands for different flux and width settings: Flux $F_1=0.38\,\rm cm^3 /s$, $F_2=0.56\,\rm cm^3 /s$, $F_3=0.65\,\rm cm^3 /s$, $F_4=0.75\,\rm cm^3 /s$, $F_5=0.85\,\rm cm^3 /s$, $F_6=0.95\,\rm cm^3 /s$ and $F_7=1.17\,\rm cm^3 /s$. Width $W_1=3 \,\rm cm$, $W_2=3.5 \,\rm cm$, $W_3=4 \,\rm cm$, $W_4=5 \,\rm cm$ and $W_5=6 \,\rm cm$.
Open circles show measurement of the bending wave speed\cite{Kim:2010vt}, and the dashed line corresponds to $\sigma = 32.7 \,\rm dyne/cm$.
\label{fig:secondfigure}}
\end{figure}

Figure \ref{fig:secondfigure} shows our measurement of the Marangoni wave speed using $v_M=v\sin\beta$ as a function of $h$.
In our experiments, $F$, $W$, and $v$ are independently varied, however the measured $\vm$ depends only on $h$.
All data points, each collected using different soap solution flux ranging $0.38 \leq F\leq 1.17 \,\rm cm^3/s$ and and channel width ranging from $3\leq W\leq 6 \,\rm cm$, collapse into a single scaling relation $\vm\propto h^{-1/2}$ in the range for $h$ spanning little less than a decade.
This clear trend that $v_M$ is a function of $h$ but not of $F$ and $W$ allows us to calculate the soap film's elasticity using $v_M = \sqrt{2E/\rho h}$\cite{Taylor:1959ta,Couder1989,Chomaz:2001us}.
The proportionality constant implies that the elasticity of our soap film is $E=22\pm1{\rm \,dyne/cm}$, being independent of $h$, $F$, and $W$, within our measurement error.

The Marangoni elastic wave is not the only wave that can propagate through a soap film; the bending wave derived by Taylor\cite{Taylor:1959ta} describes motion in which two interfaces of a film move together and moves with a speed $v_b = \sqrt{2\sigma/\rho h}$\cite{Taylor:1959ta,Couder1989,Chomaz:2001us}.
As the surface tension has the same dimensions as elasticity, the bending wave speed has the same functional dependence on the film thickness as the Marangoni wave speed.
Using previous measurements\cite{Kim:2010vt} of $\sigma\simeq32.7\,{\rm dyne/cm}$, we plotted the resulting $v_b$ in Fig. \ref{fig:secondfigure}.
The distinction of the bending wave speed from the Marangoni wave speed determined from the shock wave implies that we excite the Marangoni shock waves.


Our experimental results furnish us with new insight into physics of flowing soap films.
The soap film possesses Gibbs elasticity if sufficient time is available for perturbations in soap film concentrations to equilibrate, and possesses Marangoni elasticity otherwise.
The Gibbs elasticity $E_G =2RTc/(1+\frac{1}{2}c_bh)$\cite{Taylor:1959ta,Couder1989,Chomaz:2001us,Lucassen:1970ta,Rusanov1979} depends on $h$ and the bulk concentration $c_b$ of surfactant in equilibrium with the surface concentration $c$.
The Marangoni elasticity, $E_M = 2 R Tc$, depends on $c$ but not on $h$ or $c_b$.
We find that in the range of parameters we are able to establish the soap film, its elasticity does not depend on the soap film width or its thickness. 
The independence of the measured elasticity on film thickness implies that our flowing soap film possesses Marangoni elasticity

Furthermore, we experiment with bulk soap concentrations of 1\% and 4\% in the overhead reservoir and by changing the size of the nozzle that feeds the soap solution to the film in an attempt to influence the soap film elasticity.
We find no noticeable difference in our observation; such modifications vary the elasticity less than 4\%, falling within the margin of error.


The constant value for the Marangoni elasticity we measure and its independence on the operational parameters of the flowing soap film implies one of two possibilities: the interface is crowded with soap molecules or $\sigma \sim \sigma_0 - E (\ln c)/2$ in the parameter regime we examine.
The surface tension, $\sigma(c)$, is a function of surface soap concentration, and the accompanying Marangoni elasticity is derived using $dc/c = - dA/A$ to be $E=-2c d\sigma/dc$.
The first possibility is that in the parameter regime we explored, the soap molecules crowd the interface leading to a limiting value $c=c_\infty$.
The soap solution concentration in the overhead reservoir, and as the solution flows out and forms the film, is above the critical micellar concentration, and the surface concentration of soap molecules rapidly approaches the limiting value $c_\infty$. 
Consequently, the elasticity approaches the value of $-2c d\sigma/dc$ at $c=c_\infty$.
The alternative is that the form of $\sigma(c)$ is such that the elasticity $E$ is a constant, implying $\sigma \sim \sigma_0 - E (\ln c)/2$.
While we cannot strictly rule out the latter possibility, the former is more likely because it is the simplest explanation consistent with the observations.

\begin{figure}
\begin{centering}
\includegraphics[viewport=50bp 20bp 680bp 330bp,clip,width=3.3in]{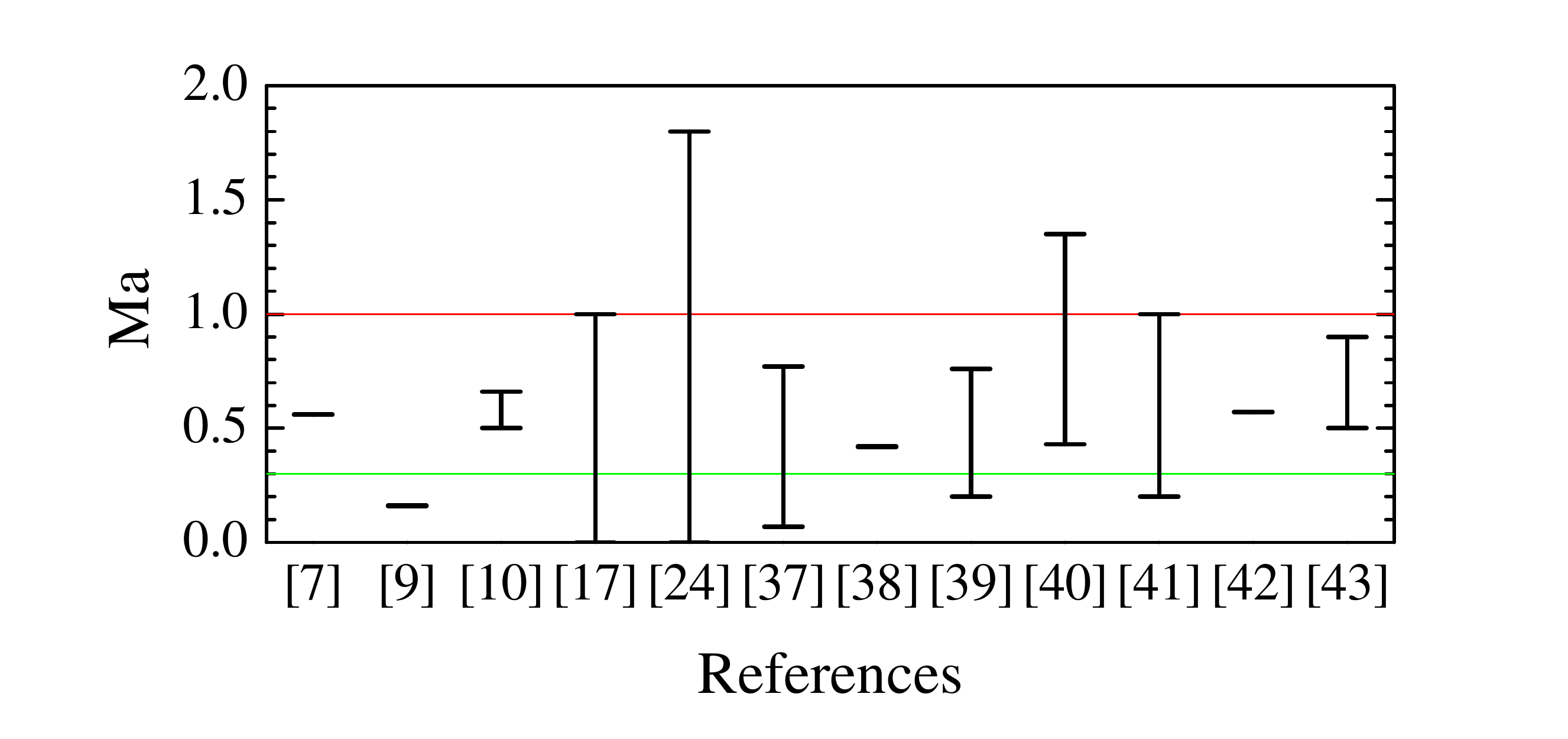}
\par\end{centering}
\caption{
Elastic Mach numbers are calculated for recent studies\cite{Vorobieff:1999wn,Kim:2015jp, Jung:2006tn,Tran:2010hn, Tran:2009vf, Alben:2002dr, Amarouchene:2004kf,Bandi:2013im,Cerbus:2013js,Jia:2009jd,Rivera:2014dw,Zhang:2000ff}. 
Using the film thickness and $E=22 \rm\, dyne/cm$, the Marangoni wave speed is calculated and used to normalized the flow speed as they are specified in the articles.
The bar graph shows the lower and upper limit of the Mach number in each studies.
Horizontal lines indicate $\Ma=0.3$ and 1.0 to guide readers.
\label{fig:lastfigure}}
\end{figure}

We postulate that our observation of the constant elasticity can be generalized, given that most soap film channel setups reported in the literature used the same soap, the same concentration, similar flow rates, and comparable dimensions for the soap film.
For such published articles which also report the film thickness, we estimate the Marangoni wave speed by assuming that the elasticity of any soap film channel is $22\rm\,dyne/cm$.
The range of Mach number is then calculated using the range of flow speed cited in each article\cite{Vorobieff:1999wn,Kim:2015jp, Jung:2006tn,Tran:2010hn, Tran:2009vf, Alben:2002dr,Amarouchene:2004kf,Bandi:2013im,Cerbus:2013js,Jia:2009jd,Rivera:2014dw,Zhang:2000ff}.
Fig. \ref{fig:lastfigure} shows the estimated range of $\Ma$ for these studies, which indicates that for the vast majority of the cases the flow is clearly of a compressible nature.
One study\cite{Tran:2009vf} recognized the compressible nature of the flow and used the Marangoni shock to estimate the compressibility, while the others do not attempt to measure the compressible character of the flow.
Our method presents a non-intrusive and low cost method for estimating the Marangoni Mach number {\it in situ} for a complete characterization of flowing soap films in future investigations.
Furthermore, the value $E=22$ dyne/cm may be used to determine the Marangoni Mach number without any experimentation.

To summarize, we provide an experimental method for {\it in situ} measurement of the Marangoni wave speed.
In our method, we artificially generate oblique shocks in soap film flows by introducing an obstruction, and determine the Marangoni Mach number by measuring the shock angle.
Our measurements show that the Marangoni wave speed depends on the film thickness, and the elasticity is constant for films in our range of experiments independent of film thickness, width, flow rate, or the bulk concentration of surfactants. 
We suspect that the elasticity is constant in our soap films because soap concentration is higher than the critical micelle concentration.
Considering that it is hard to establish a soap film using a dilute soap solution,  we suspect that the reported value $22 \,\rm dyne/cm$ of the elasticity must be universal for all soap films using the same recipe.



\bibliographystyle{apsrev4-1}
\bibliography{soapfilm2015}

\end{document}